# Popularity based Bandwidth Allocation for Video Broadcast/Multicast over Wireless Networks

Mostafa Zaman Chowdhury[a,£], Young-Il Kim[b], Won Ryu[b], and Yeong Min Jang[a,*]
[a] Department of Electronics Engineering, Kookmin University, Korea
[b] Electronics and Telecommunications Research Institute (ETRI), Korea
E-mail: [£]mzceee@yahoo.com, [*]yjang@kookmin.ac.kr

*Abstract*— Recently, video broadcast/multicast over wireless networks has created a significant interest in the field of wireless communication. However, the wireless resources have limitations to broadcast/multicast many video sessions at the same time with the best quality. Hence, during the video transmission through wireless networks, it is very important to make the best utilization of the limited bandwidth. When the system bandwidth is not sufficient to allocate the demanded bandwidth for all of the active broadcasting/multicasting video sessions, instead of allocating equal bandwidth to each of them, our proposed scheme allocates bandwidth per video session based on popularity of the video program. Using the mathematical and simulation analyses, we show that the proposed scheme maximizes average user satisfaction level. The simulation results also indicate that a large number of subscribers can receive a significantly improved quality of video. To improve the video quality for large number of subscribers, the only tradeoff is that a very few subscribers receive slightly degraded video quality.

## I. Introduction

During last couple of years, a notable development of broadband wireless access networks has been observed. Mobile WiMAX is a typical example of an emerging wireless network system. The Mobile WiMAX is capable of providing high data rate with flexible Quality of Service (QoS) mechanisms, making the support of Mobile TV very attractive. The fast deployment of broadband wireless networks has raised expectation of real-time video services in mobile environments. However, limited bandwidth is a challenge for supporting high data rate video services through the wireless link. Therefore, transmission of videos through the wireless link using broadcasting or multicasting technique has become very popular approach compared to the unicast approach.

Scalable video technique [1]-[3] is used for the variable bit rate video broadcast/multicast over wireless networks. This technique utilizes multiple layering. Each of the layers improves spatial, temporal, or visual quality of the rendered video to the user [1]. Base layer or the highest priority layer guarantees the minimum quality of a video stream. Whereas the addition of enhanced layers or low priority layers improves the video quality. The number of layers for a video session (program) and the bandwidth per layer can be manipulated dynamically. Thus, to broadcast/multicast videos through a wireless environment, layered transmission is an effective approach for supporting heterogeneous receivers with varying bandwidth requirements [3]. Hence, if the system bandwidth is not sufficient to allocate the demanded bandwidth for all of the active broadcasting/multicasting video sessions, it is possible to allocate higher bandwidth for the popular video session compared to less popular one. In this paper, we address two important problems in video broadcast/multicast over wireless networks: 1) maximizing the average user satisfaction level and 2) the best utilization of the network bandwidth.

Due to the limited data rates of wireless networks, it is not possible to provide the best quality for the entire active broadcasting/multicasting video sessions. Hence, equal bandwidth allocation for all of the broadcasting/multicasting video sessions is an easy and simple way. The service qualities of all broadcasting/multicasting video sessions are equally degraded when the total wireless bandwidth is not sufficient to provide the maximum demanded bandwidth to all. Instead of allocating equal bandwidths to all of the sessions during an insufficient bandwidth condition, our proposed scheme efficiently allocates the total system bandwidth among them in such a way that higher bandwidth is allocated to the video session of higher popularity. Thus, the average user satisfaction level is increased significantly. However, a minimum quality for the lowest popular broadcasting/multicasting video session is guaranteed by assigning a minimum amount of bandwidth.

The rest of this paper is organized as follows. Section II shows the system model for the proposed bandwidth allocation scheme. The detail calculation of user satisfaction level is also shown in this section. The performances of the proposed scheme are verified in Section III. Finally, conclusions are drawn in the last section.

## II. Bandwidth Allocation Scheme

In the past few years, there has been extensive works on video broadcast/multicast over wireless networks [1]-[6]. Yu Wang *et al.* [2] presented a variable bit rate allocation for the broadcasting of scalable video over wireless networks. The authors in this paper proposed the variable bit rate allocation for the base layer as well as for the enhanced layers. Jiangchuan Liu *et al.* [3] proposed the layering in multisession video broadcasting. Jen-Wen Ding *et al.* [4] proposed the spectrum-based bandwidth allocation algorithm for layered video streams over wireless broadcast channels.

---

[*]Corresponding author of the paper

An easy and straightforward approach is that all of the active broadcasting/multicasting video sessions share the total system bandwidth equally. However, such approach is not sensible. Because a popular video program attracting a large number of subscribers should be allocated with more bandwidth compared to the less popular one, if allocation of total demanded bandwidth is not possible. Our proposed scheme allocates the bandwidth per broadcasting/multicasting video session based on popularity of the session.

Let the total system bandwidth capacity and the total number of active broadcasting/multicasting video sessions are $C$ and $M$, respectively. $\beta_{max}$ and $\beta_{min}$ are, respectively, the maximum allocated bandwidth and the minimum allocated bandwidth for each of the active broadcasting/multicasting video sessions. So, the system can provide minimum $\lfloor C/\beta_{max} \rfloor$ and maximum $\lfloor C/\beta_{min} \rfloor$ numbers of video sessions simultaneously. The allocated bandwidth for each of the active sessions in the equally shared bandwidth allocation scheme is:

$$\beta = \begin{cases} \beta_{max}, & \beta_{max}M \leq C \\ \dfrac{C}{M}, & \beta_{max}M > C \end{cases} \quad (1)$$

Larger amount of allocated bandwidth for a video session makes the chance of increasing the number of enhanced layers and thus improving the video quality for that session. User satisfaction level depends on received video quality. Therefore, we assume that user satisfaction level is directly proportional to allocated bandwidth for a video session. User satisfaction level becomes maximum (equal to 1) when the demanded bandwidth ($\beta_{max}$) is allocated for a broadcasting/multicasting video session. The satisfaction level of a user in the equally shared bandwidth allocation scheme can be written as:

$$S_L = \begin{cases} 1, & \beta_{max}M \leq C \\ \dfrac{C}{\beta_{max}M}, & \beta_{max}M > C \end{cases} \quad (2)$$

Our proposed scheme allocates different amount of bandwidths for different broadcasting/multicasting video sessions based on popularity of video program. However, the maximum allocated bandwidth to a broadcasting/multicasting video session is $\beta_{max}$ and the minimum allocated bandwidth to a broadcasting/multicasting video session is $\beta_{min}$. Where $\beta_{min}$ ensures the minimum quality of a video session. An active broadcasting/multicasting video session is ranked based on the number of users currently watching the program on that session.

The most popular broadcasting/multicasting video session (program) is ranked as 1. Where the lowest popular one is ranked as $M$. The numbers of active users for different broadcasting/multicasting video sessions are related as:

$$\left.\begin{array}{l} K_1 \geq K_2 \geq \cdots \geq K_m \geq \cdots \geq K_M \\ K = K_1 + K_2 + \cdots + K_m + \cdots + K_M \end{array}\right\} \quad (3)$$

where $K_m$ is the number of users watching the *m-th* video program. $m=1$ indicates that program which is being watched by the maximum number of users. Whereas $m=M$ indicates that with the minimum users. $K$ is the total number of active users in the system.

Fig. 1 shows the basic concepts of bandwidth allocation per broadcasting/multicasting video session by the equally shared and the proposed popularity based bandwidth allocation schemes when the system bandwidth is not sufficient to allocate $\beta_{max}$ for each of the active broadcasting/multicasting video sessions. Fig. 1(a) shows that an equal bandwidth $\beta$ is allocated to each of the broadcasting/multicasting video sessions by the equally shared bandwidth allocation scheme. On the other hand, Fig. 1(b) shows that the same bandwidth is not allocated to each of the active video sessions by the proposed popularity based bandwidth allocation scheme. Maximum bandwidth $\beta_1$ is allocated to the broadcasting/multicasting video session #1 which is enjoyed by the maximum number of subscribers. On the other hand, minimum bandwidth $\beta_M$ is allocated to the broadcasting/multicasting video session #M which is received by the minimum number of subscribers.

Bandwidth $\beta_{max}$ is allocated for each of the broadcasting/multicasting video sessions whenever $\beta_{max}M \leq C$. However, if $\beta_{max}M > C$, then the allocated bandwidth $\beta_m$ for *m-th* broadcasting/multicasting video session in the proposed popularity based bandwidth allocation scheme is calculated by the following procedures,

$$X_m = \begin{cases} 0, & \left(aK_m + \sum\limits_{j=1}^{m-1} X_j\right) \leq \beta_{diff} \\ \dfrac{aK_m + \sum\limits_{j=1}^{m-1} X_j - \beta_{diff}}{M-m}, & \left(aK_m + \sum\limits_{j=1}^{m-1} X_j\right) > \beta_{diff} \end{cases} \quad (4)$$

where $a = \dfrac{M}{K}\left(\dfrac{C}{M} - \beta_{min}\right)$ and $\beta_{diff} = \beta_{max} - \beta_{min}$

$$\beta_m = \begin{cases} \beta_{max}, & \left(aK_m + \sum\limits_{j=1}^{m-1} X_j\right) \geq \beta_{diff} \\ \beta_{min} + aK_m + \sum\limits_{j=1}^{m-1} X_j, & \left(aK_m + \sum\limits_{j=1}^{m-1} X_j\right) < \beta_{diff} \end{cases} \quad (5)$$

Hence, the allocated bandwidths of the active broadcasting/multicasting video sessions for the proposed scheme are related as:

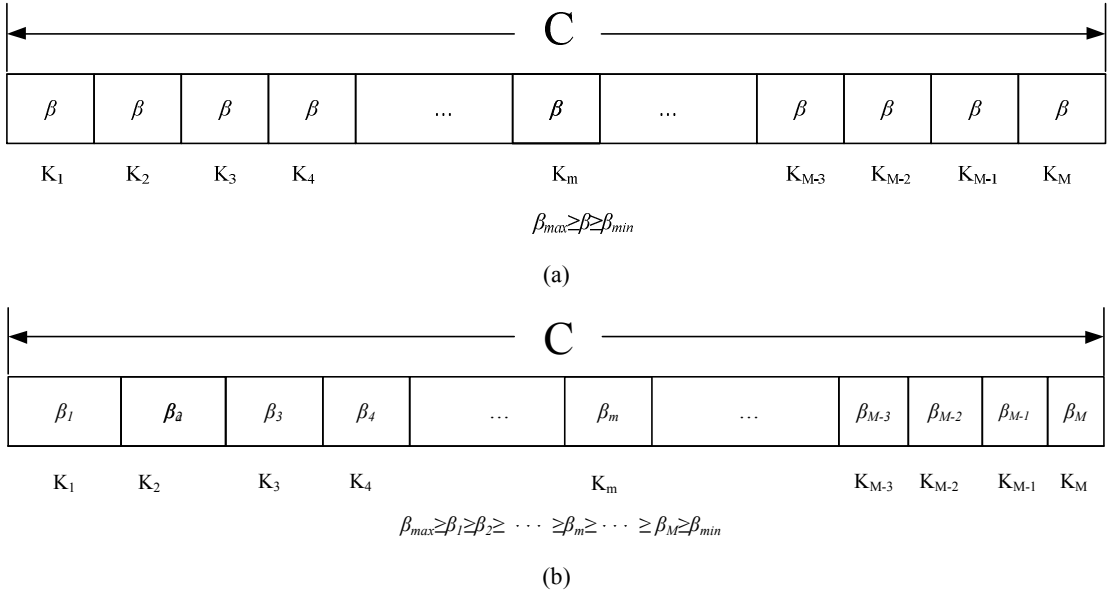

**Fig. 1.** An example of bandwidth allocation when the system bandwidth is not sufficient to allocate $\beta_{max}$ for each of the broadcasting/multicasting video sessions (a) equal allocated bandwidths to all of the broadcasting/multicasting video sessions by the equally shared bandwidth allocation scheme, (b) different allocated bandwidths to different broadcasting/multicasting video sessions by the proposed popularity based bandwidth allocation scheme.

$$\left.\begin{array}{l} \beta_1 \geq \beta_2 \geq \cdots \geq \beta_m \geq \cdots \geq \beta_M \\ \beta_1 \geq \dfrac{C}{M} \\ \beta_M \leq \dfrac{C}{M} \end{array}\right\} \quad (6)$$

Whenever the number of users for an active video session or the total number of active video sessions is changed, the allocated bandwidth for each of the active broadcasting/multicasting video sessions is also dynamically changed. As a consequence, the number of enhanced layers per session and the allocated bandwidth per enhanced layer may also be changed. It can be mentioned that a receiver cannot subscribe to a fraction of a layer.

In our proposed scheme, the satisfaction level of the users who are connected with the *m-th* broadcasting/multicasting video session is:

$$S_{L(m)} = \begin{cases} 1, & \beta_{max}M \leq C \\ \dfrac{\beta_m}{\beta_{max}}, & \beta_{max}M > C \end{cases} \quad (7)$$

The average user satisfaction level for the proposed scheme is calculated as:

$$S_{L(av)} = \begin{cases} 1, & \beta_{max}M \leq C \\ \dfrac{\sum_{m=1}^{M} S_{L(m)} K_m}{K}, & \beta_{max}M > C \end{cases} \quad (8)$$

where $SL_{(av)}$ is the average user satisfaction level for the proposed scheme considering all the active users in the system.

The relation between the average user satisfaction levels for the proposed popularity based bandwidth allocation scheme and the equally shared bandwidth allocation scheme can be written as:

$$\left.\begin{array}{ll} S_{L(av)} = S_L = 1, & \beta_{max}M \leq C \\ S_{L(av)} = S_L, & K_1 = K_M \\ S_{L(av)} > S_L, & K_1 \neq K_M \text{ and } \beta_{max}M > C \end{array}\right\} \quad (9)$$

### III. Performance Evaluation

In this section, we verified performance of the proposed scheme using simulation results. We assume bandwidth capacity $C$ = 30 Mbps; for a broadcasting/multicasting video session maximum allocated bandwidth $\beta_{max}$ = 2 Mbps; and minimum allocated bandwidth for a video session $\beta_{min}$ = 0.6 Mbps. We consider random number of active users for per video session while the total number of active users in the system is always 200.

Firstly, we verify the improvement of average user satisfaction level for our proposed scheme compared to the equally shared bandwidth allocation scheme. Fig. 2 shows that the proposed scheme provides much better average user satisfaction level compared to the equally shared bandwidth allocation scheme. The user satisfaction level decreases with the increase of active video sessions due to the limited bandwidth capacity of the network.

Fig. 3 shows a comparison between the numbers of users to whom the video quality is improved and the users to whom it is degraded in the proposed scheme compared to the equally shared bandwidth allocation scheme. Fig. 3 indicates that huge number of users can enjoy improved video quality. To improve the video quality for these large number of users, the only adjustment is that a very few users receive slightly degraded video quality. Hence, a large number of users enjoy the significantly improved video quality.

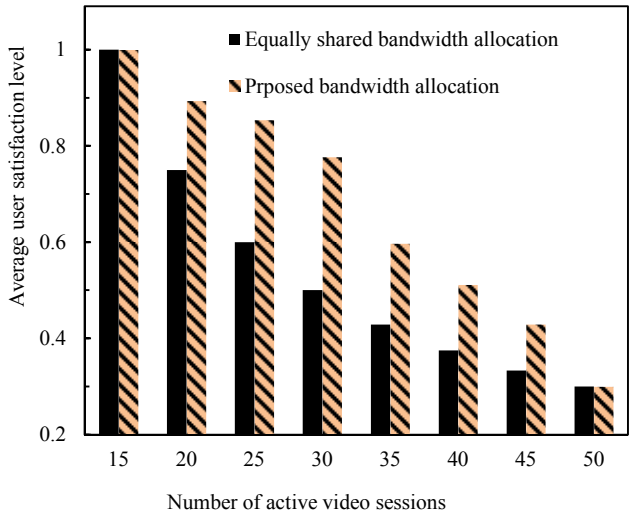

**Fig. 2.** A comparison of the average user satisfaction levels for various numbers of active video sessions.

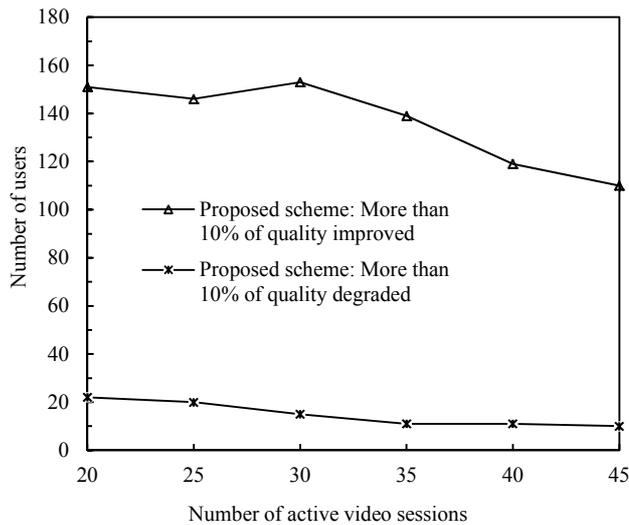

**Fig. 3.** A Comparison to show the number of users to whom video quality is improved or degraded in the proposed scheme with respect to the equally shared bandwidth allocation scheme.

The results in Figs. 2 and 3 show that our proposed popularity based bandwidth allocation scheme is able to improve average user satisfaction level. The proposed scheme is even more effective when large number of users watch the program of a common broadcasting/multicasting video session.

## VI. Conclusions

This paper proposes an efficient bandwidth allocation scheme for the real-time video broadcast/multicast over wireless networks. The proposed scheme allocates bandwidth for each of the broadcasting/multicasting video sessions based on the importance of the sessions during the lack of bandwidth. We compare the proposed scheme with the equally shared bandwidth allocation scheme to show the performance improvement. This paper also demonstrates how the popularity of a video session affects the bandwidth allocation. Simulation results indicate that the proposed bandwidth allocation scheme is very effective for video broadcast/multicast over the wireless networks.

## Acknowledgement

This work was also supported by Electronics and Telecommunications Research Institute (ETRI), Korea.